# Mobile solutions for clinical surveillance and evaluation in infancy - General Movement Apps

Peter B Marschik[1,2,3]*, Amanda KL Kwong[4,5,6]*, Nelson Silva[3]*, Joy E Olsen[4,5] Martin Schulte-Rüther[1], Sven Bölte[2,7,8], Maria Örtqvist[9,10], Abbey Eeles[4,5,6], Luise Poustka[1], Christa Einspieler[3], Karin Nielsen-Saines[11]*, Dajie Zhang[1,3]*~, Alicia J Spittle[4,5,6]*

1. Child and Adolescent Psychiatry and Psychotherapy, University Medical Center Göttingen, 37075 Göttingen, Germany and Leibniz-ScienceCampus Primate Cognition, 37075 Göttingen, Germany
2. Center of Neurodevelopmental Disorders (KIND), Centre for Psychiatry Research; Department of Women's and Children's Health, Karolinska Institutet, 11330 Stockholm, Sweden
3. iDN – interdisciplinary Developmental Neuroscience, Division of Phoniatrics, Medical University of Graz, 8036 Graz, Austria
4. Murdoch Children's Research Institute, Parkville, 3052 Victoria, Australia
5. The Royal Women's Hospital, Parkville, 3052 Victoria, Australia
6. Department of Physiotherapy, The University of Melbourne, Parkville, 3052 Victoria, Australia
7. Curtin Autism Research Group, Curtin School of Allied Health, Curtin University, 6102 Perth, Western Australia
8. Child and Adolescent Psychiatry, Stockholm Health Care Services, Region Stockholm, 11861 Stockholm, Sweden
9. Department of Women's and Children's Health, Neonatal Research Unit, Karolinska Institutet, 11330 Stockholm, Sweden
10. Functional Area Occupational Therapy & Physiotherapy, Allied Health Professionals Function, Karolinska University Hospital, 11330 Stockholm, Sweden
11. Division of Pediatric Infectious Diseases, David Geffen UCLA School of Medicine, 90095 Los Angeles, USA

~ please address correspondence to: Dajie.marschik@med.uni-goettingen.de

*these authors share first or senior-authorship

**Keywords**

**Abstract**

The Prechtl General Movements Assessment (GMA) has become a clinician and researcher tool-box for evaluating neurodevelopment in early infancy. Given it involves observation of infant movements from video recordings, utilising smartphone applications to obtain these recordings seems like the natural progression for the field. In this review, we look back on the development of apps for acquiring general movement videos, describe the application and research studies of available apps, and discuss future directions of mobile solutions and their usability in research and clinical practice. We emphasise the importance of understanding the background that has led to these developments while introducing new technologies, including the barriers and facilitators along the pathway. The *GMApp* and *Baby Moves* App were the first ones developed to increase accessibility of the GMA, with two further apps, *NeuroMotion* and *InMotion*, designed since. The Baby Moves app has been applied most frequently. For the mobile future of GMA, we advocate collaboration to boost the field's progression and to reduce research waste. We propose future collaborative solutions including standardisation of cross-sites data collection, adaption to local context and privacy laws, employment of user feedback, and sustainable IT structures enabling continuous software updating.

**Introduction**

**How it all came about – from Prechtl's first observations of general movements at the bench side to the use of smartphone recordings**

In the late 1980s, the first, systematic comparison of spontaneous movements in preterm and term infants indicated a qualitative, but not a quantitative difference in early movement patterns pointing towards neurodivergence [1]. These observations marked a starting point for the development of the general movements assessment (GMA [2]), a method for evaluating the integrity of the young nervous system through the assessment of overt spontaneous motor behaviour. The GMA is a clinical reasoning approach based on visual Gestalt perception of typical vs. atypical movements in the entire body, hence the term *general* movements (GMs) [2-4].

From the 9th week postmenstrual age to approximately 20 weeks' post-term age, foetuses/infants show a distinct repertoire of endogenously generated (i.e., not triggered by sensory input) movement patterns such as startles, GMs, breathing movements, yawning, and sucking (e.g., [4,5]). Normal GMs present themselves in a variable sequence of neck, trunk, leg and arm movements, with gradual beginnings and endings and of variable intensity, force and speed [3,6]. Before term age, GMs are commonly referred to as foetal or preterm GMs, whereas GMs observed between term age and approximately 6 to 8 weeks post-term age are called writhing movements (WMs). Normal WMs can last between seconds and several minutes. WMs gradually disappear during the second month post-term, and a new pattern of GMs – known as fidgety movements (FMs) – emerges [2,7]. Normal FMs are small movements of moderate speed with variable acceleration of the neck, trunk and limbs in all directions. They are continually observable during active wakefulness and are highly predictive of normal neurodevelopment [2,7]. Even though there is growing evidence of sex/gender effects in evolving neurodiversity [8], endogenously generated neurofunctions like GMs seem to be sex/gender-general (e.g., [5,9,10]).

For analysing phenomena with complex appearances, such as GMs, Gestalt perception is a powerful tool but depends on the observer's skills and is vulnerable to distracting environmental influences. The reliability of the GMA has been examined in several studies, suggesting that trained observers consistently achieve an excellent inter-observer agreement ranging between 89 and 98%, or an average Cohen's kappa of 0.88 [1,11]. Since the development of GMA, there have been numerous studies on the inter-rater reliability for GMA, based on traditional recording methods [12-19]. However, recently, GMA researchers (i) are optimizing existing recording techniques through technological developments, (ii) amending detailed clinical protocols and assessments based on visual Gestalt perception (e.g., [20-22]), and (iii) developing computer-based approaches to analyse GMs and changing the research and clinical use of GMA (e.g., [23-37]). Even though we observe an increase in 'tech-based' studies on GMs overall and the results from computer-based GMA appear promising, they cannot yet replace human visual Gestalt perception in the evaluation of GMs [24,25,33,37].

In recent years, technological innovation and improvement have led to an increase in attempts to refine recording techniques or develop new tools (i – iii above) adapted to different research needs or clinical practice. New inventions have evolved to meet the need for adaptation to changing recording settings (e.g., in the clinics vs. at home), to enable coverage of wide geographic areas, to allow non-GMA specialists (parents and healthcare providers) to gather valid data sets, and to create big data for the analyses of infant motor behaviour. The application of smartphone-based solutions has been at the forefront, since these devices allow for recordings in different settings, in the clinic or remote and help accurate documentation of the timing of data acquisition. Our group of authors was the first to develop smartphone-based solutions for GMA. With this article we aim to (1) provide an overview of general movement app developments; (2) describe the use and research studies of available apps; and (3) discuss future directions of mobile solutions and their usability in research and clinical practice.

**Materials and Methods**
**Search Engines, Inclusion Criteria, Paper Extraction and Selection**

For the literature review, we systematically searched for publications related to mobile applications to assess infant spontaneous motor functions, or general movements (GMs) more specifically. We included recording tools used by caregivers or healthcare professionals to collect remote video recordings, upload video recordings to a remote server, and/or provide feedback to GMA experts. Any study that used a smartphone app for GMA data collection was included in this review. We included protocol papers as well as empirical studies written in English and published in peer-reviewed journals or conference proceedings and preprints. Fifteen databases and research networks were searched in August 2022 and again in March 2023; PubMed, WoS, Science Direct, arXiv, PLOS, SpringerLink, Nature, Frontiers, Elsevier, Research Gate, SCOPUS, and Google Scholar. In addition to these sources, we also searched in Google for personal webpages, blogs, forums, patents, GitHub, Apple-, Google- (Android) and Windows- (Mobile) app stores, and performed an ancestral research of published papers for additional studies. For any paper where an app was first described, a citing literature search was performed on Medline OVID and screened for inclusion. Search terms and Boolean Operators were as follows: smartphone OR tablet OR mobile OR remote software OR system OR android OR app* OR apple OR iOS OR iPhone OR eHealth OR e-Health OR mHealth OR m-Health OR patent* OR general movement* OR GMA OR Prechtl OR spontaneous motor* OR neuromotor OR infant motor OR infant move*. The search resulted in a total of 33 records. In the screening step, we first deleted all duplicates and checked titles and abstracts against inclusion criteria (AK, NS). We then removed studies not related to mobile applications for GMA.

**Results**

According to our search and screening procedures, we identified a total of six records reporting on mobile solutions for GMA (Table 1) and additional eleven to make use of the developed apps. These articles relate to 4 different Apps: *Baby Moves* [38], *GMApp* [32], *NeuroMotion* [39] and *In-Motion* [35]. The basic functionality is similar for all related apps, recording GMs and sending videos to a remote server. While all provide instructional guidelines for obtaining video recordings of infant spontaneous movements for later assessment of GMs and upload to the server, there are specifics pertinent to each tool (Tables 1, 2). Three apps were developed for home-recordings by parents, whereas *GMApp* was primarily designed for use by healthcare professionals or in healthcare environments. Each app has a different feedback process. For example, the *GMApp* provides an *OEOC feedback loop* (Observer - Expert - Observer – Caregiver; Figure 2). This loop represents a daily clinical application scenario: an observer records an infant with the app initiating the *OEOC cascade*; the expert conducts the GMA and informs the observer about the outcome. If indicated, the caregivers will be provided with an intervention plan and/or referred to a specialist. Whereas for the *Baby Moves* App there is purposefully no feedback to the family via the app, rather the process of reporting back the outcome of the assessment and next steps is individualised to the local research setting. Parent-user feedback from the *NeuroMotion* app noted that combining app use with face-to-face visits would be preferable in some instances [39].

[insert Table 1 about here]

In all the applications, users take videos at defined time points according to GMA [3] and receive specific recording aids such as date reminders, instructions for proper recording, baby silhouette to guarantee full-body coverage, etc. (Table 2).

[insert Table 2 about here]

The *Baby Moves* App is the most studied to date and has been used in several populations including extremely preterm, moderate to late preterm, healthy term born and hypoxic-ischaemic encephalopathy, some studies being related to the Victorian Infant Collaborative Study 2016-2017 cohort [38,40-49]. Following the initial studies to validate the use of the *Baby Moves* App, including parent feedback, the app was updated to version 2.0.1, which improved user experience and number of quality videos uploaded for assessment [50]. *GMApp* was used in studies on a neurotypical cohort, a preterm cohort and a cohort reporting about prenatal Zika virus infection [7,9,21,32]. The apps have already been adapted to a variety of languages (e.g., *Baby Moves*: English, Spanish, Italian, Arabic; *GMApp*: English, German, Portuguese; *NeuroMotion*: English and Swedish; *InMotion*: English, Norwegian). The three apps targeting parent-users, *Baby Moves*, *NeuroMotion* and *InMotion*, have all reported positive parent feedback about app usability [35,39,44]. Inter-rater reliability based on video recordings from smartphone apps are similar to those obtained using traditional methods. Raters scoring videos recorded via the *Baby Moves* app achieved an intra-class correlation coefficient of 0.77 (95% CI 0.75-0.80) for fidgety movements [51]. For videos obtained via the *NeuroMotion* app, k-alpha statistic agreement ranged from 0.48-0.72 with all raters achieving moderate-almost perfect agreement [39].

**Discussion**

Since the late 1990s, GMA has become an increasingly and widely used assessment tool in clinical and research settings. With eHealth and mHealth (mobile health) developments, smartphone apps for recording and processing GMs have been integrated into research studies in several countries.

**From different parts of the world to a global endeavour and a joint goal**

Not surprisingly, two groups started initiatives to develop smartphone-based solutions to aid recording and assessment of GMs at the same time (Spittle et al. [38] supported by the Cerebral Palsy Alliance and the Murdoch Children's Research Institute; Marschik et al. 2017 [32] supported by a Grand Challenges Explorations Grant 2015 of the Bill and Melinda Gates Foundation). In the spirit of collaboration, lead investigators of both GMApp and *Baby Moves* came together to discuss collaboration after the first developmental steps of the individual apps. Around that time, another group of GMA experts also started their endeavour to develop a similar app [35,52]. Even though initial software development was conducted by distinct groups across different continents, it soon became evident that all experts faced similar obstacles and had comparable goals. While the *Baby Moves* app was the first one to be used by parents, initially in a specific geographic setting, *GMApp* was originally built for use by healthcare providers in Low- and Middle-Income Countries (LMIC). Given the *Baby Moves* app and *GMApp* targeted different users they shared knowledge on app development and processes, however, maintained separate apps. The *NeuroMotion* app was developed later and used the experience of these prior apps. App development consistently aimed towards the same goal, which was to facilitate broader GMA implementation through the use of mHealth technology.

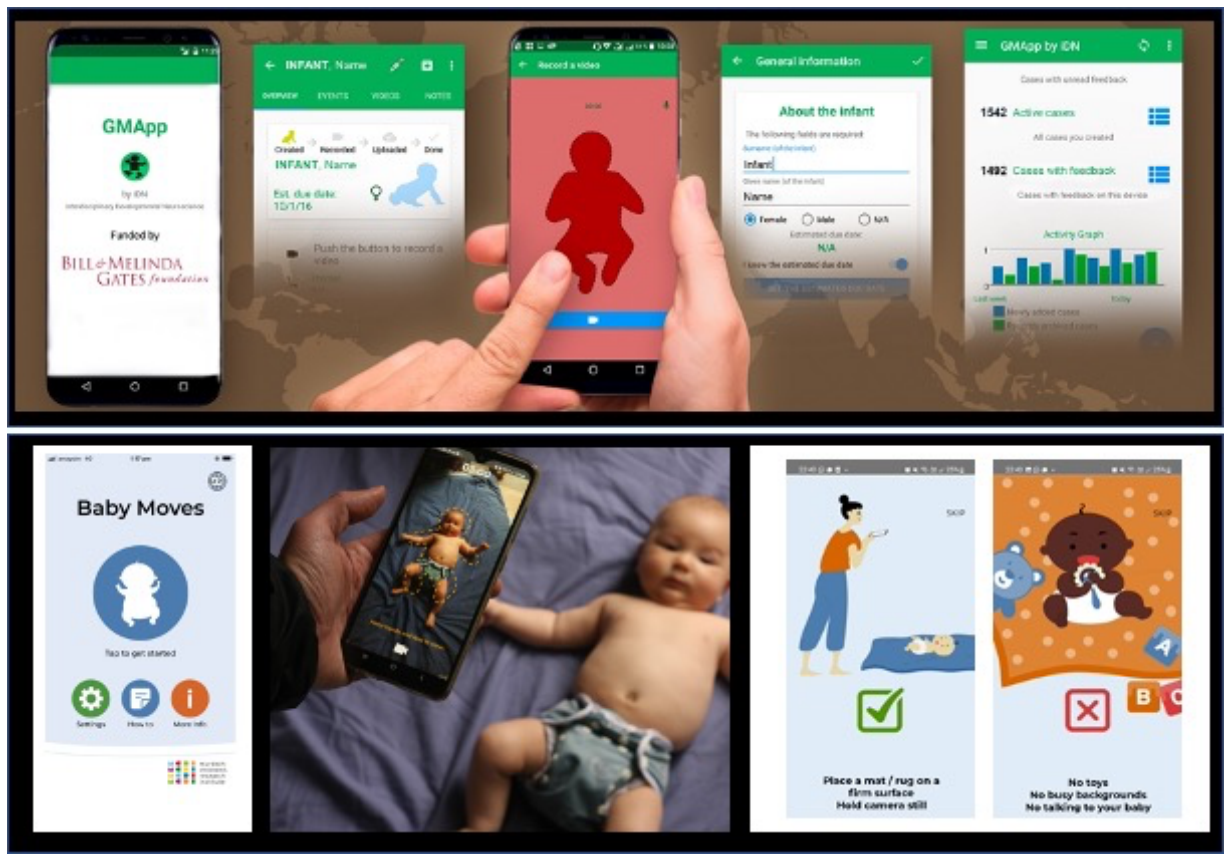

**Figure 1.** The first General Movement apps: 1a. Screenshots of *GMApp* including baby silhouette and statistics; 1b. Baby silhouette of *Baby Moves* (a video demonstration of the latest version can be accessed via [50]).

**User experience**

An important aspect of smartphone apps is the user experience and app usability. As outlined in specific articles about mHealth quality criteria, tools need to be evaluated for accuracy, efficiency, effectiveness and usability [53]. There are some common features in the GM apps, including reminder notifications for recording GMs within specific periods, baby silhouette (Figure 1), and set recording length time (all apps had certified tutors or at least experienced GMA experts in the development phase), which enhanced usability. The *GMApp*, for example, has undergone intensive evaluations (Grand Challenges Explorations Project Report, Bill and Melinda Gates Foundation; program details at https://gcgh.grandchallenges.org/grant/gmapp-developing-brain-and-developing-world-hand) from GMA experts and non-experts (N=17) in four iteration circles leading to an average satisfaction score of 4.43/5 concerning applicability, learnability, and completeness of the app. Instructions for recording GMs enable better quality recording of GMs for scoring. The *Baby Moves* app has been updated to incorporate user feedback, and to include step by step instructions and infographics, while studies have used different instructional guides and modes (written, video) to increase the likelihood of scorable GMs being recorded [44]

Research on smartphone apps focused on parent-recorded videos has evaluated the parent's experience of recording and uploading their child's videos, using structured questionnaires [39,44]. In these studies, the majority of respondents found the apps easy to use while uploading videos was simple and safe. Similarly, healthcare providers using the *GMApp* also reported that it was intuitive and easy to use. In regards to video quality using an app, for the *GMApp,* a study of smartphone-GMA vs. classic GMA recordings was conducted and asked eight experts to evaluate the differences (Grand Challenges Explorations Project Report, Bill and Melinda Gates Foundation; program details at https://gcgh.grandchallenges.org/grant/gmapp-developing-brain-and-developing-world-hand). They were not able to distinguish whether a video was made by an HD camcorder or the GMApp.

The GMA provides information about an infant's neurodevelopment, and as such, recording of GMs via an app is only one aspect of the assessment process for infants and their families. Following the performance of GMA, feedback to families about their infant's assessment with potential recommendations for further neurodevelopmental assessments should be made if needed. Some apps currently have additional features to aid recording, such as optimizing data transfer volume or a fussing detector (i.e. an algorithm that automatically stops the recording when the baby becomes fussy or starts to cry), while guaranteeing that videos appropriate for future expert analysis are generated (Grand Challenges Explorations Project Report, Bill and Melinda Gates Foundation; program details at https://gcgh.grandchallenges.org/grant/gmapp-developing-brain-and-developing-world-hand). For the fussing detector we classified 2830 vocalizations (cry vs. fuss vs. no cry), and analysed their acoustic features. In a pilot assessment, we achieved, using linear kernel support vector machines, an unweighted recall of 77,2% [54].

**Challenges and Future directions**

Although smartphone apps are particularly useful for observational assessments such as GMA, there are a number of challenges to be dealt with. Ongoing improvements in user instructions and technical recording aids have improved the quality and ease of app-based GMs recording. We suggest that current developers should join forces and use current apps as a blueprint for an integrated framework that is agile for future GMs assessment infrastructure (Figure 2).

Future applications should consider integrated GMA within a clinical pathway or research workflow. This would allow video data to relay seamlessly to GMA assessors in a timely manner that links with patient and family feedback within clinical management. Researchers using the *GMApp* piloted such an approach with the *OEOC feedback loop* (Observer – Expert – Observer – Caregiver), thus enabling healthcare providers not trained in GMA to coordinate such assessments. Furthermore, such feedback loops could also be used for individualized training of GMA for healthcare providers and to develop strategies for more efficient, large-scale assessments. Specifically, for the case of early screening for cerebral palsy, the GMA is recommended alongside brain neuroimaging and physical neurological examination [55]. Integrating GMA data with other early assessments is key to maximizing clinical efficiency and assessment accuracy.

Recent developments and research into machine learning approaches suggest that fully automated AI-supported assessments to assist clinical reasoning may be feasible in the future [24]. However, careful refinement of algorithms with large cohorts [33] and efficient involvement of human GMA experts ("human-in-the-loop"[56]) will be needed to translate these approaches into clinical practice and test their utility to support clinical decisions.

Joint efforts will be needed in the era of "big data" to collect sufficiently large and diverse datasets that prevent bias and allow for objective evaluation of classification algorithms. Standardized app-based assessments in combination with a solid server infrastructure, data management and data standardization are pivotal in this respect. Integrating a smooth user experience, broad functionality, universal applicability and standardized data collection requires considerable resources in terms of technical, administrative, and legal support. Both an "app" (mobile front-end), and a server back-end infrastructure needs to be implemented, including databases, data storage and data transfer management. The current fast-paced technical development of operating systems, mobile platforms and frameworks requires continuous updating and maintenance. Additionally, country-specific legal requirements with respect to privacy and data security need to be balanced carefully with ease of use, organizational policies and research requirements. Furthermore, when databases/cohorts from different research groups and healthcare providers are collated, integrated, and shared, steering policies and data access policies need to be developed and implemented.

The effort for such a joint integrated approach would be worthwhile: It will be easier to acquire sufficient data for ML models that may aid clinical decision, or provide direct feedback and enhancements during recording with the app (Figure 3; e.g., real-time user support, such as video overlay hints, camera adjustment/positioning and video recording quality checks, instant infant body pose estimation/detection). Furthermore, useful features such as face anonymization [33], remote consultation services, or integration of new technical developments only need to be implemented once, and could be used directly across the community. Exciting current technical advancements include, for example, depth sensors in mobile devices (e.g., LIDAR sensors), augmented reality hardware and frameworks, and dedicated mobile AI processing hardware and frameworks. In addition, the possibilities for data sharing, collaboration, would increase exponentially, and data standardization would greatly enhance data re-usability according to the FAIR principles [57]. A central database provides the possibility to include rich metadata, optimal conditions for dataset publication, and access policies for re-use of dataset by qualified researchers.

The general movements apps are a use case for smartphone-based research methods. They have immense potential to increase our understanding of infant development, delineate trajectories, understand causalities and assess the effects of interventions. They are not substituting but rather complementing lab- or clinic-based research and are utilized in different fields spanning from infant neuromotor functions (use case) to mental health (e.g., [58]).

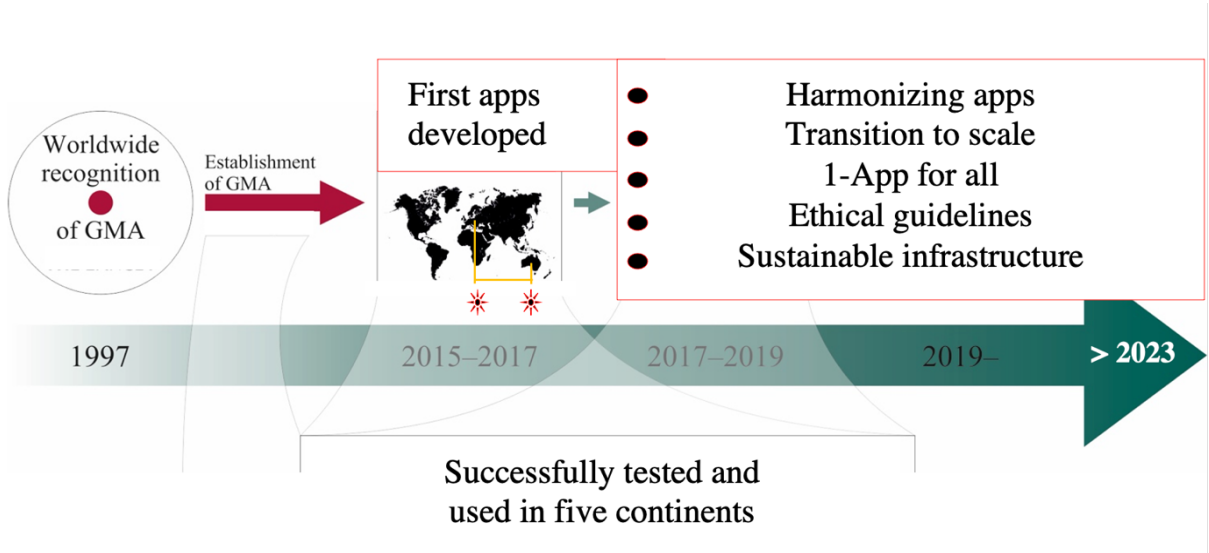

**Figure 2.** From the beginnings of GMA to the first general movement smartphone apps to scaling-up and global use.

- **What have we learned?**
- Smartphone apps for recording infant motor functions can be successfully used by most parents and healthcare providers.
- Lower maternal education, limited language skills and reliance on government financial support was related to poorer engagement.
- Reliability between GMA assessed using traditionally GM methodology and using smartphones applications.
- The use of smartphone apps has enabled collection of large data sets in research trials across varied clinical/diagnostic populations.

- **Considerations for the future!**
- Collaboration to reduce duplication and progress the field
- Apps that can be altered to the local context
- Ethical regulations
- Data protection guidelines
- Continuous software updating
- Sustainable IT structures and privacy preserving ML/AI capabilities
- Involvement of specialized organizations (e.g., GM-Trust, CP-alliance)

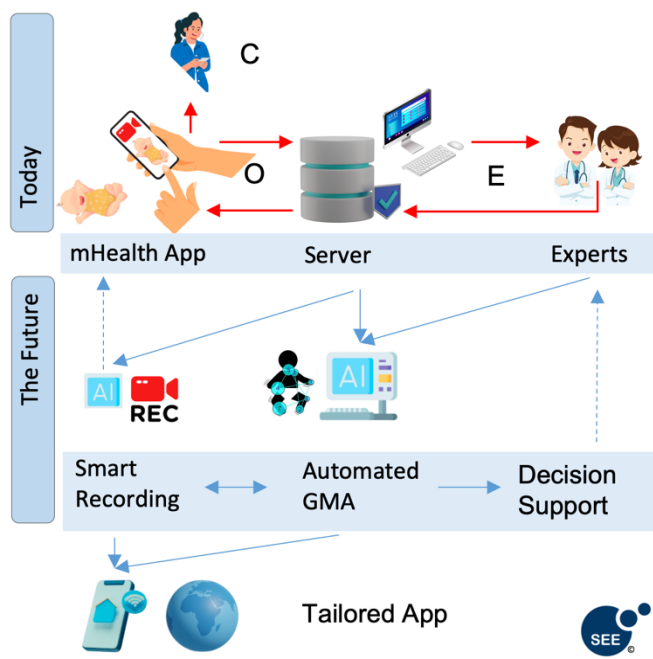

**Figure 3.** Frontend-backend pipeline of all described apps including OEOC feedback loop of GMApp. Perspective of integrated AI to aid recording and automated assessment of GMs for decision support. Key: C, caregiver; O, observer; E, expert.

**How would Heinz Prechtl comment on these developments?**

Heinz Prechtl [59], who was open to technological advances and at the same time sceptical of their utility before they proved to be valuable and reliable, would have supported the idea of standardized data collection allowing for comparability, open data science and scaling-up. While unsure about economic gains companies would achieve to enable investment in such developments, he would have promoted this endeavour for use on a scientific platform, and so do all authors of this article: "Thoroughly documenting development in small samples is extremely valuable and important. Scaling it up and not forgetting to continue working thoroughly sure is too. As with everything else, not quantity but also the quality matters." (He actually never said this precisely, but he certainly could have).

**Conclusion**

To conclude, as telemedicine, eHealth and mHealth development is fast and increasingly embedded in our healthcare systems, only a joint endeavour bringing these developments together while adapting them to local systems will improve global usage and sustainability. This can bring forth an essential contribution to families with children who are at an elevated likelihood for adverse or neurodiverse outcomes while also supporting multi-centre or large-scale research activities aiming at deciphering early infant development.

Even though there have been a number of recent developments in this field, all with similar scientific and clinical goals, only a joint venture and a sustainable consortium will substantially bring this field forward. We need to synergize knowledge and expertise of GMA experts, developmental scientists, paediatricians, physiotherapists, computer scientists, software engineers, app developers, cyber security experts, data managers and last but not least, law experts to guarantee data protection and confidentiality.

What looks like a "simple app" is a complex undertaking; to create a smartphone-to-server solution while guaranteeing world-wide coverage. This endeavour requires a concerted effort to create a sustainable tool and an adequate environment for the mobile-future of GMA.

**Conflicts of Interest**

Professor Peter Marschik is President of the GM-Trust. Professor Christa Einspieler and Professor Alicia Spittle are licensed tutors with the General Movements Trust. Peter Marschik developed together with Dr. Dajie Zhang and Christa Einspieler the *GMApp*. Professor Alicia Spittle, Dr. Abbey Eeles, Dr. Joy Olsen and Dr. Amanda Kwong contributed to the development of the *Baby Moves* app.

**Acknowledgements**

We dedicate the idea of new ways for GMA to our late friend and mentor Heinz Prechtl, knowing he would have appreciated these efforts to unify and synergize our efforts aiming to help the youngest, who need us the most. We also thank Magdalena Krieber-Tomantschger for helping with the literature research and working together with PBM, DZ, and CE on the FWF supported grant (KLI 811); the Leibniz Foundation (LSC Audacity Award); and the Volkswagen Foundation (IDENTIFIED); PBM, LP, and MSR are supported by the DFG (grant numbers: 456967546, SCHU2493/5-1). DZ was supported by the DFG, SFB1528 – Pj C03; *GMApp* development was supported by the Bill and Melinda Gates Foundation (OPP112887). The *Baby Moves* App has been supported by the Cerebral Palsy Alliance, Murdoch Children's Research Institute, and the National Health and Research Council of Australia.

Table 1: Characteristics of publications on mobile solutions referring to the four different apps for GMA.

| Authors | App | Study/ article type | Participants |
|---|---|---|---|
| Spittle et al., 2016 | *Baby Moves* | Protocol, app description | Extremely preterm infants, term control infants |
| Kwong et al., 2018 | *Baby Moves* | Prospective cohort | Extremely preterm infants (n = 204), term control infants (n = 212) |
| Elliott et al., 2021 | *Baby Moves* | Protocol | General population prospective cohort (aim: 3000 infants) |
| Marschik et al., 2017 | *GMApp* | Protocol, project-related website, app description | Neurotypical cohort, convenience sample of preterm infants (aim: 50 infants longitudinally in biweekly intervals) |
| Svensson et al., 2021 | *NeuroMotion* | Prospective cohort, app description | Infants at heightened risk for adverse neurological outcome (n = 52; 95 parents) |
| Adde et al., 2021 | *In-Motion* | Prospective cohort, app description | Infants at high-risk for cerebral palsy (n = 86) |

Table 2: App description and functionality.

| App / Name | Original app description | Platform | Users | Reminders / Functionality | Video upload | Recording aids |
|---|---|---|---|---|---|---|
| *Baby Moves* | Spittle et al., 2016; | iOS, Android | Parents | Filming start reminder (push notification) | Server upload to REDCap | • Instruction guidelines<br>• Countdown timer<br>• Baby silhouette |
| *GMApp* | Marschik et al., 2017* | iOS, Android | Healthcare professionals | Filming start date reminder function | Server upload / cloud system as *GMApp* has been | • Instruction guidelines<br>• Baby silhouette<br>• Jitter detector |

| | | | | *OEOC- feedback loop* / video feedback to healthcare providers and caregivers<br><br>Case management for the assessment of multiple infants<br><br>Statistics and overview about patients | used in different continents and countries | - Fussing detector<br>- Brightness detector<br>- Minimum recording-length regulator<br>- Automated recording stop and upload |
|---|---|---|---|---|---|---|
| *NeuroMotion* | Svensson et al., 2021 | iOS, Android | Parents | Filming start date recording reminder (push notification) | Server upload to REDCap | - Instruction guidelines<br>- Screen filter to capture infant's whole body<br>- 2-3 minute timer |
| *In-Motion-App* | Adde et al., 2021 | iOS, Android | Parents | 1 week prior reminder<br>Filming start date reminder<br>Visual recording timeline | Server upload (server at St. Olavs Hospital, Norway) | - Instruction guidelines in video format (2m 47s duration)<br>- 3-minute auto-recording stop |

*in this publication, the GMApp was introduced and details were available on a project related website which was deactivated in 2020. Details about GMApp can be accessed through the corresponding author of this article